\documentclass[twocolumn,showpacs,preprintnumbers,amsmath,amssymb,aps,pre,10pt]{revtex4-1}
\pdfoutput=1
\usepackage{graphicx,color}
\usepackage{rotating}

\begin{document}

\title{How fast does the clock of Finance run? -- A time-definition
  enforcing scale invariance and quantifying overnights}

\author{Michele Caraglio}
\email{caraglio@pd.infn.it}
\affiliation{Dipartimento di Fisica e Astronomia Universit\`a di
  Padova and sezione CNISM, Via Marzolo 8, I-35131 Padova, Italy} 
\author{Fulvio Baldovin}
\email{baldovin@pd.infn.it}
\affiliation{Dipartimento di Fisica e Astronomia Universit\`a di
  Padova, sezione INFN, and sezione CNISM, Via Marzolo 8, I-35131
  Padova, Italy} 
\author{Attilio L. Stella}
\email{stella@pd.infn.it}
\affiliation{Dipartimento di Fisica e Astronomia Universit\`a di
  Padova, sezione INFN, and sezione CNISM, Via Marzolo 8, I-35131
  Padova, Italy}

\date{\today}

\begin{abstract}
    A symmetry-guided definition of time may enhance and simplify
    the analysis of historical series with recurrent patterns and
    seasonalities.  By enforcing simple-scaling and stationarity of
    the distributions of returns, we identify a successful protocol of
    time definition in Finance.  The essential structure of the
    stochastic process underlying the series can thus be analyzed
    within a most parsimonious symmetry scheme in which multiscaling
    is reduced in the quest of a time scale additive and
    independent of moment-order in the distribution of returns.
    At the same time, duration
    of periods in which markets remain inactive are properly
    quantified by the novel clock, and the corresponding (e.g.,
    overnight) returns are consistently taken into account for
    financial applications.
\end{abstract}

\pacs{
}

\maketitle

\section{Introduction}
Relativity theory~\cite{Wald1984} provides a remarkable example of the
construction of a nontrivial time scale on the basis of a symmetry principle;
namely, the proper time of
a moving body from the requirement of invariance of the laws of Nature for different
observers. 
This example, in which symmetry lays at the
very foundation of the physical theory, may suggest to
look for other contexts in which symmetries 
could help in defining useful notions of time.
Nonstationary time series are very frequently encountered
in fields like
Meteorology~\cite{Ausloos1999},
Seismology~\cite{Corral2004,Corral2006},
Physiology~\cite{Goldberger1995,Stanley2001,Hausdorff1995},
Economy and Finance~\cite{Liu1997,Galluccio1997,Bassler2007}.
The nonstationary mark limits the applicability of statistical methods
in the quest for the process characterizing the series~\cite{Tsay2002},
and may convey spurious effects in the detection of
correlations in the time series~\cite{Peng1994,Kantelhardt2002}.
Depending on the quality of the stochastic process,
different strategies can be applied to simplify the signals and to
recover stationarity properties.
For instance, various detrending 
procedures have been proposed~\cite{Peng1994,Kantelhardt2002},
although in the general case
their success is not always guaranteed~\cite{Chen2002,Bryce2012}.
Other approaches~\cite{Dacorogna1996,Dacorogna2001,Zumbach2002,Corral2004}
are instead based on a reassessment of the time stamp in terms of which the
series is recorded. In particular, in the case of seismology
it has been recently
advanced~\cite{Corral2004,Corral2006} that 
the distribution of interoccurrence times between earthquakes 
becomes scale-invariant under a proper redefinition
of time; the latter renders interoccurrencies stationary on the basis of the Omori's
aftershock law~\cite{Omori1984}. 

Following inspiration provided by the example of relativity,
within the context of time series affected by cyclic nonstationary 
behavior~\cite{Gardner2006} we show here that one can promote
the approximate fulfillment of a simple scaling symmetry to become an
operative criterion for the definition of a novel time,
which we call financial scaling time (FST).
Our focus is on Finance, a field in which the very definition 
of time constitutes a long-standing problem, while the 
application of symmetry principles lacks a strong tradition.
As one should expect, the symmetry enforcement leads to
a distinct simplification of the analysis.
Specifically we show that financial returns, which are 
nonstationary in physical time, become so in FST.
Within the financial context, multiscaling
has been largely reported as a ``stylized fact'' and has inspired
the construction of multifractal
models~\cite{Frisch1997,Mandelbrot1998,Calvet2002,Borland2005,DiMatteo2007,Bacry2008}.
Although we prove in Appendix A that our time
redefinition cannot completely remove multiscaling effects,
we concurrently show that they are in fact reduced in FST; consistently,
the probability density functions (PDF) of returns
satisfy simple scaling to a good approximation in a 
rather wide range of time scales.
The FST also offers a natural quantification of the time elapsing 
during markets closures, so that the associated returns can be
consistently taken into account in financial applications.
Albeit focused on Finance, our methodology could be of broader
physical interest -- for instance in the analysis of interoccurrence
times~\cite{Corral2004,Corral2006,Bogachev2007} --
exemplifying how a symmetry-based time transformation may render
a stochastic process stationary, as required in 
many statistical approaches~\cite{Chicheportiche2014,Chicheportiche2017}.  

A clock properly adjusted to financial activity is not easy to
identify.  Financial markets have to cope with human daily routines
throughout the world and bursts and doldrums occur at various time
scales. In addition, during nights, weekends, and festivities,
transactions stop in most cases. These interruptions, at the end of
which assets prices turn out to have changed anyhow, together with the
recurrent patterns in the activity (seasonalities), immediately signal
the inadequacy of ``natural'' physical time as the appropriate one in
terms of which to describe the stochastic evolution of financial
markets.  Besides physical time, different alternatives have been
studied~\cite{Clark1973,Dacorogna2001,Ane1999,Stanley2000,Jensen2004,gillemot2006},
including trading time, volume time, and tick-by-tick time.  A time
definition (theta-time) has also been put forward with the specific
intent of getting rid of the seasonalities in financial time
series~\cite{Dacorogna1996,Dacorogna2001,Zumbach2002}.  Theta-time is
designed to record the progress of market activity through the
increase of the volatility as measured by the average absolute
return.
For the latter a power-law behavior is assumed as a function
of physical time, but no discussion of the univocity of this time 
definition and of its effectiveness in enforcing stationarity of
the return PDF is made.
In most theoretical studies, 
inactive market periods are cut from the analyzed
dataset and the corresponding returns ignored~\cite{Dacorogna2001}.
However, the practice of
ignoring inactive periods destroys the correspondence between the sum
of returns and the real asset price, and as such it is thus not
suitable to many financial applications.  Alternatively, one could
assign an arbitrary time interval to overnight and similar returns,
but this implies altering the time scaling properties of their PDFs.
Hence, the appropriate duration to be ascribed to overnight and
similar returns remains an open issue.
In general, there is also little focus on the requirement that
increments over intervals of equal span should be identically
distributed in order to make statistical sampling (e.g.,
sliding-window) applicable. 

Here we show that a solution to these problems is naturally suggested
by the existence of an approximate symmetry which appears when the PDFs
increments over different intervals are compared.
Indeed, from days to several weeks, the empirical
financial returns' PDFs are found to be approximately scale-invariant, in
a sense familiar from the Physics of critical phenomena~\cite{Baldovin2007}.
Defining financial returns $r$ over the time span $\Delta t$ 
(here $\Delta t$ is an integer number of days)  as the
log-difference of the asset price value $s$,
$r\equiv\ln s(t)-\ln s(t-\Delta t)$, this means that the empirical PDFs of $r$ over 
different $\Delta t$'s
can be approximately collapsed onto each other in force of the scaling law
\begin{equation}
p(r,\Delta t) \simeq \frac{1}{\Delta t^H} \; g \left( \frac{r}{\Delta t^H} \right),
\label{eq_scaling}
\end{equation}
where $H$ is called the Hurst exponent~\cite{Hurst1951},
and $g$, which is not Gaussian, is a scaling function.
It turns out that $H$  is close to $1/2$ for assets
of developed markets~\cite{Galluccio1997,DiMatteo2005}.

As discussed in Appendix A, a natural way of defining time
is that of referring to a specific $q$-th order moment of the return PDF's,
$\mathbb{E}[|r|^q]$ $(q>0)$,
assuming that the time interval is directly measured by this moment.
This definition promotes the stationarity of the PDF over intervals
of equal duration in the new time scale.
However, we shown in Appendix A that
only if in the novel time the PDF satisfies
a form of simple scaling like in Eq.~(\ref{eq_scaling}) with $H=1/2$,
the time definition is independent of the chosen moment order and the novel
time is additive.
On the contrary, a simple scaling with $H \neq 1/2$ would prevent additivity and,
most important,
in the case of multiscaling, when $H$ depends on the moment order $q$,
the time scales are different for different orders
and there is no more a unique additive time in terms of which one
can describe the returns aggregation process.

The strict validity of Eq.~\eqref{eq_scaling} with $H=1/2$ in physical time
would thus establish an univoque correspondence between returns' PDFs
and time intervals durations $\Delta t$'s and we could say that physical
time provides the time scale we are in quest of.
However,
Eq.~\eqref{eq_scaling} definitely does not hold
at the intraday time scales, where nonstationarities 
affect the PDFs for returns defined over intervals with  
equal physical time duration~\cite{Galluccio1997,Bassler2007,Allez2011},
and $H$ can be very different from $1/2$ in selected time windows~\cite{Bassler2007,Baldovin2015}.
Even in the interday domain multiscaling
effects~\cite{Calvet2002,Borland2005,DiMatteo2007} 
and a slow crossover to Gaussianity~\cite{RamaCont2001} 
prevent a full realization
of the collapses implied by Eq.~(\ref{eq_scaling}).

In view of the difficulties involved in the definition of time
in Finance, specifying the returns' PDFs in terms of a single, 
appropriate time scale remains a basic goal worth pursuing, 
also at the cost of relying on a symmetry scheme which can only 
be satisfied approximately.
Here we show that if defined with respect to a suitable 
function of physical time -- namely, the FST --
returns' PDFs 
become almost identical for intervals of the same FST duration.
Moreover, when rescaled through an Hurst exponent equal to $1/2$,
satisfactory collapses for PDFs related to different FST spans are
exhibited.
Thus enforcement of simple scaling allows to define a univoque
time scale and to guarantee an optimal degree of stationarity for
the whole return PDF, not just for one of its moments.
Our approach relies on empirical data only and as
such can be considered model-free;
our results hold within a window ranging from few minutes to several days in physical time, hence bridging the intraday and interday regimes.

\section{Basic ideas and constraints}
Our basic ansatz is that, once expressed in terms of FST, 
Eq.~\eqref{eq_scaling} turns into a scaling law with $H=1/2$:
\begin{equation}
p(r, \Delta \tau)=\frac{1}{\Delta \tau^{1/2}}
\,\;g\left(\frac{r}{\Delta \tau^{1/2}}\right) \, ,
\label{eq_scaling2}
\end{equation}
where $\Delta \tau>0$
represents the FST duration of an interval and $g$ is
a suitable scaling function.
As clarified in Appendix A, Eq.~(\ref{eq_scaling2}) is not consistent with established multiscaling properties of financial time series~\cite{DiMatteo2007,Wang2008}.
However, our results clearly demonstrate that Eq.~(\ref{eq_scaling2}) is {\it approximately valid} within a window ranging from few minutes to several days, provided that the proper time scale, namely the FST, is suitably defined in order to enforce at best such scaling symmetry.

The FST construction exploits the fact that, 
in spite of the manifest nonstationarities at the
intraday level~\cite{Galluccio1997,Dacorogna2001,Bassler2007,Allez2011}, the process of return aggregation
appears compatible 
with the assumption of one-day
ciclostationarity~\cite{Gardner2006}.
In other words, the evolution of
the aggregate return from the opening during each market day 
can be regarded as a realization of the same stochastic 
process~\cite{Dacorogna2001,Bassler2007,Baldovin2015}, and empirical averages can be
taken over correspondent time windows within different days.
The FST construction proceeds as follows: on the physical time axis we first
operate a partition into intervals consistent with the one-day
periodicity of the process of return formation.
Intervals are arbitrary, except for the fact that
their extrema should not fall within the periods of market closure,
and that their duration should be above a lower cutoff (typically of
the order of a few minutes, as we discuss below). 
It is
convenient, but not necessary, to set the morning opening of the
market as the lower extremum of the first
partition interval, and to let the last 
interval of each day coincide with the full overnight (or weekend) closure.
In the example treated below, we further choose to assign the same
physical time span to all remaining intraday intervals.
Intervals exceeding the duration of one day are then constructed as
the union of a suitable number of
intraday intervals pertaining to different days.

Our purpose is to associate an appropriate FST-duration 
to each interval with extrema belonging to the
partition. Of course, as discussed in Appendix A, if we call
$\Delta \tau_1$  and $\Delta \tau_2$ the FST duration of
any two contiguous intervals, we require the span $\Delta \tau$ of 
their union to satisfy the measure property
\begin{equation}
\Delta \tau = \Delta \tau_1 +\Delta \tau_2.
\label{eq_measure}
\end{equation} 
This allows us to consistently map the physical-time partition onto 
the $\tau$-axis.
So, a $t$-axis partition possesses a 
$\tau$-axis partition image, and vice versa.
Eq.~\eqref{eq_scaling2} together with Eq.~\eqref{eq_measure}
immediately impose an important constraint
which allows to easily locate a time lower limit for the applicability
of the proposed scheme.
Calling $r_1$ and $r_2$ the returns over the contiguous intervals of duration  
$\Delta \tau_1$  and $\Delta \tau_2$, respectively, we necessarily have 
$\mathbb{E}\left[(r_1+r_2)^2\right]=(\Delta\tau_1+\Delta\tau_2)\int\mathrm{d}x\,g(x)\,x^2 = \mathbb{E}\left[r_1^2\right] + \mathbb{E}\left[r_2^2\right]$;
whence
\begin{equation}
\mathbb{E}\left[r_1\,r_2\right]=0,
\label{eq_uncorrelation}
\end{equation}
where $\mathbb{E}$ denotes ciclostationary averages in the sense specified above.
This linear uncorrelation, which does not imply independence~\cite{RamaCont2001},
corresponds to the martingale property for the stochastic process
defining the price formation~\cite{Bouchaud2003} and 
is generally valid if the market is efficient and the interval
duration is above a few minutes (in physical time)~\cite{RamaCont2001,Bouchaud2003}.
This is the origin of the lower cutoff to be assumed in the duration
of the partition intervals.

\section{Results}
For convenience, 
we take as a reference sample the empirical PDF of 
a full day (opening-to-opening) return, 
and associate to this time interval the unit of FST:
$1\;\mathrm{day}=1\;\mathrm{fst}$.
To assess and quantify the validity of
Eq.~\eqref{eq_scaling2},
we extensively apply the Kolmogorov-Smirnov (KS) 
two-sample test~\cite{DeGroot2010,Darling1957}:
the scaling factor $\Delta\tau$ guaranteeing through
Eq.~\eqref{eq_scaling2} the best possible data-collapse with respect to the
reference sample PDF, identifies the duration in $\mathrm{FST}$ units for the chosen
returns' interval.

Enforcing the validity of Eq.~(\ref{eq_scaling2}) gives
a clear practical advantage over the option of defining FST with 
reference to, e.g., the second moment alone.
In fact with our choice the FST turns out to be
consistent with the simple scaling
ansatz on a wider window (see Table~\ref{tab_DeltaTau}).
This is probably due to the fact that emphasizing the scaling 
constraints of a single moment risks to overlook the role
played by other moments in the determination of a satisfactory
time scale.

In the Methods Section we describe in detail the implementation of the
above ideas to the S\&P500 index,
recorded at 1-minute frequency between 9:40 and 16:00 from September 1985
to June 2013~\cite{DataSetStatement}. 
For this specific dataset (including entries of thirty years ago) a
duration $\overline{\Delta t}$
over which contiguous returns can be considered
linearly uncorrelated corresponds to 20-minutes (See Methods for
details). Due to the impact of information technology on trading
practice,
the more recent the data are considered, the lower can be
put this threshold~\cite{Bouchaud2003}.

A partition of the physical-time axis is conveniently identified
as 
$\{t_{l,m}\}_{l=0,1,\ldots;\;m=0,1,\ldots,m_{\rm max}}$, 
where $l$ labels the day
after the one whose opening has been chosen as the origin
($t_{0,0}$),
and $m$ singles out the time instant within day $l$
(\mbox{$t_{0,0}<t_{0,1}<\ldots<t_{0,m_{\rm max}}<t_{1,0}<t_{1,1}<\ldots$}).
In view of the previous discussion, we require 
\begin{equation}
t_{l,m+1}-t_{l,m}\geq\overline{\Delta t}
\qquad\forall l\;\textrm{and for }m>0,
\end{equation}
and $t_{l,0}$ and $t_{l,m_{\rm max}}$ to correspond to the
daily opening and closure times, respectively.
Below, a generic return at time $t$ [$\tau$] 
over the time-scale $\Delta t$ [$\Delta\tau$]
will be indicated as 
\mbox{$r_{\Delta t}(t)\equiv \ln s(t) - \ln s(t-\Delta t)$} 
[\mbox{$r_{\Delta\tau}(\tau)\equiv \ln s(\tau) - \ln s(\tau-\Delta \tau)$}], or,
in terms of a partition $\{t_{l,m}\}$,
\mbox{$r_{l,m}^{l',m'} \equiv \ln s(t_{l',m'}) - \ln s(t_{l,m})$},
with $t_{l',m'}>t_{l,m}$.

\begin{figure}[tbp]
\centering \includegraphics[width=0.46 \textwidth]{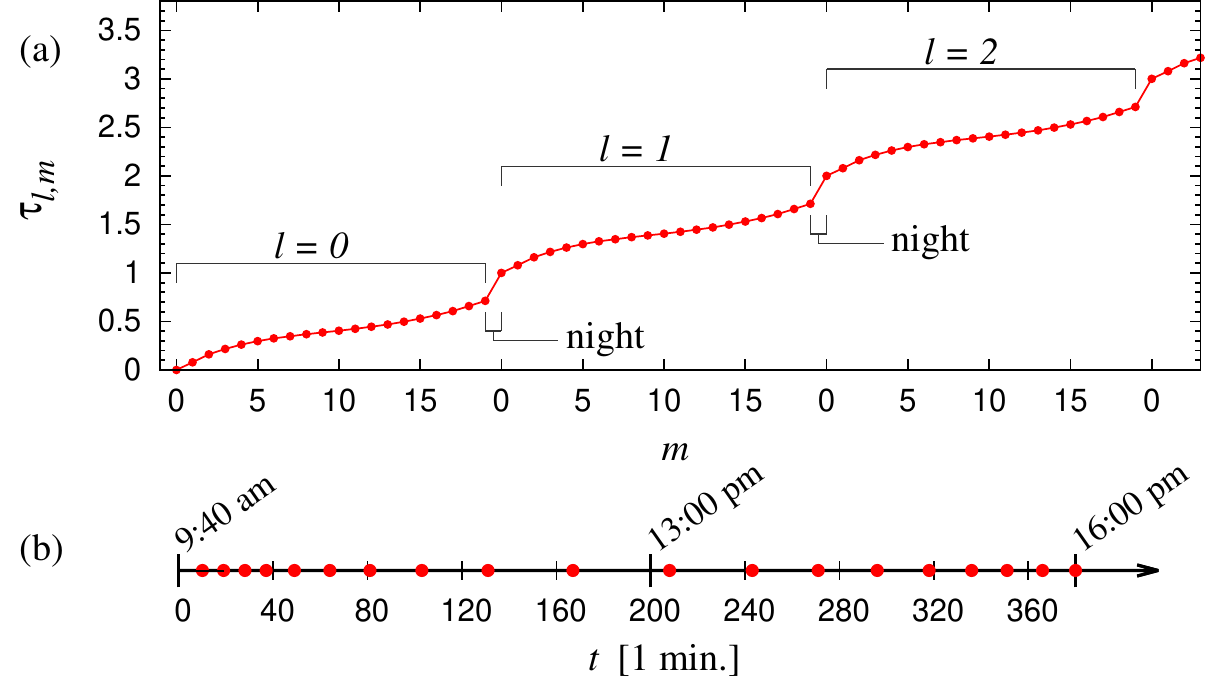} 
\caption{ 
  FST vs. physical time. In (a) the integer $l$ labels the day after the
  chosen reference one and the integer $m$ the 20-minute multiple after
  the opening time. 
  In (b) instants occurring at multiples of $\Delta\tau=0.037\;{\rm fst}$
  from the opening (red circles) 
  are plotted on the physical time axis.
}
\label{fig_time}
\end{figure}

Independently of the day $l$, the KS two-sample test enables us to determine the duration in FST,
$\Delta\tau_{m}$ (with $m>0$), of each of the $m_{\rm max}$ partition intervals occurring during a
day. For instance, the first 20-minute interval of the day corresponds to 
$\Delta\tau_{1}=0.08\;\mathrm{fst}$, whereas twenty minutes at half
and at the end of the day amount to 
$\Delta\tau_{11}=0.02\;\mathrm{fst}$ and
$\Delta\tau_{19}=0.05\;\mathrm{fst}$, respectively
(See Methods for details). 
Considering the close-to-open return
\mbox{$r_{m_{\rm max},0}^{0,1}$},
we can also establish the duration of an overnight interval in the FST:
$\Delta\tau_{\rm night}=0.29\;\mathrm{fst}$.
Interestingly, the duration of overweekends (Friday-close-to-Monday-open)
is about the same. 
Consistently with the requirement in Eq.~\eqref{eq_measure}, the FST scale 
is hence constructed as
\begin{equation}
\label{eq_TimeDefinition}
\tau_{l,m} \equiv \sum_{n=1}^{m} \Delta \tau_{n}
+  l \, \left(  \sum_{n=1}^{m_{\rm max}} \Delta \tau_{n} 
+ \Delta \tau_{\rm night} \right) \; .
\end{equation}
Notice that, assuming Eq.~\eqref{eq_scaling2}, returns over different intervals with the
same $\Delta\tau$ are by construction identically distributed.

\begin{figure*}[tbp]
\centering \includegraphics[width=0.8 \textwidth]{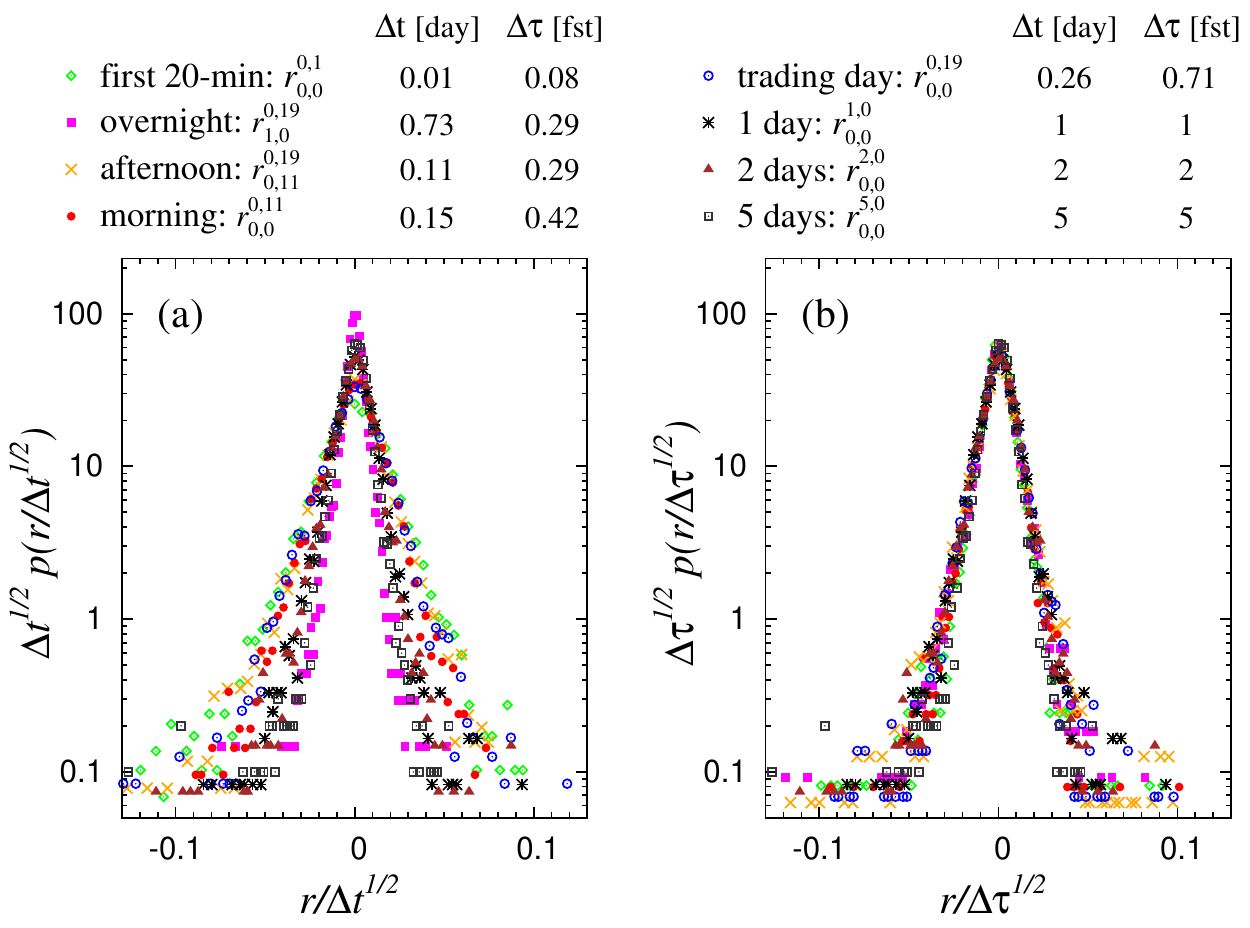}
\caption{
  Empirical PDFs rescaled according to Eq.~\eqref{eq_scaling} in (a), and
  Eq.~\eqref{eq_scaling2} in (b).
  While in the physical clock the rescaled PDFs do not 
  satisfy a scaling symmetry since they do not collapse onto each
  other (a), when proper time is adopted (the FST) the
  scaling symmetry nicely emerges as rather satisfactory data-collapse (b).
}
\label{fig_Collapse}
\end{figure*}

In Fig.~\ref{fig_time}a we outline the identification of the FST for an
equally-spaced partition, 
$t_{l,m+1}-t_{l,m}=\overline{\Delta t}=20\;{\rm min}$, $m_{\rm max}=19$.
The result is obtained by averaging over all possible days
in the dataset.
Fig.~\ref{fig_time}a highlights the non-linear character of the FST
vs. the physical one.
If, reversely, one plots equally-spaced 
contiguous $\Delta\tau$-intervals as a function of the physical time
(Fig.~\ref{fig_time}b), it becomes evident that the FST runs faster at
the beginning and at the end of the day, and slower at noon,
New York time. 

A qualitative inspection about how the FST definition emphasizes
the simple scaling properties of the empirical PDFs is offered in
Fig.~\ref{fig_Collapse} (Tables~\ref{tab_KS1},~\ref{tab_DeltaTau} 
in the Method Section quantify these results). 
The comparison of Fig.~\ref{fig_Collapse}a with
Fig.~\ref{fig_Collapse}b makes evident that only in terms of the FST
the data-collapse implied by  
Eq.~\eqref{eq_scaling2} can be assumed to hold
from 20 min up to several days. 
As a side note, we observe that both the overnight and the afternoon duration  
lasts less than the morning
interval.

\begin{figure*}[tbp]
\centering \includegraphics[width=0.86 \textwidth]{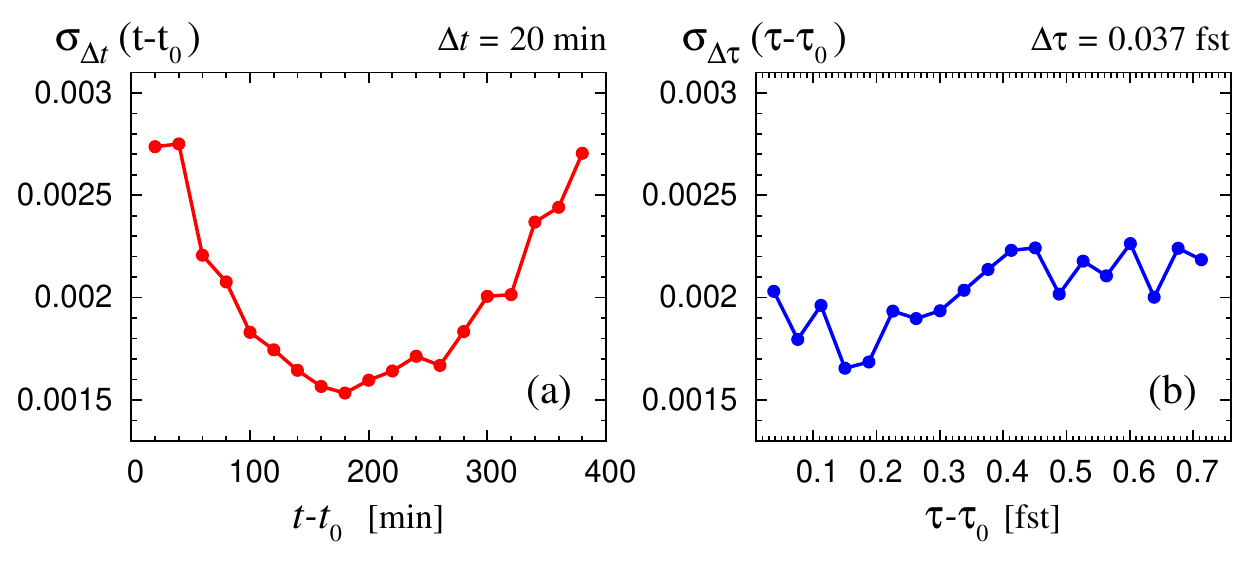}
\caption{Intraday volatility. Each market day is assumed to be an
  independent realization of the same stochastic
  process~\cite{Bassler2007,Baldovin2015}. Volatility over the scale
  $\Delta t$ (a) and $\Delta\tau$ (b) is plotted as a function of the
  time elapsing from the market opening ($t_0$ and $\tau_0$,
  respectively).
  While in physical time volatility is markedly
  nonstationary (a), apart from fluctuations it becomes stationary
  in FST (b). 
}
\label{fig_TimeComparison}
\end{figure*}

Volatility is a central quantity in financial practice, assessing the intensity of
market fluctuations. It may be defined as
\mbox{$\sigma_{\Delta t}(t)\equiv{\mathbb{E}\left[\left|r_{\Delta t}(t)\right|\right]}$}
-- in FST:
\mbox{$\sigma_{\Delta\tau}(\tau)\equiv{\mathbb{E}\left[\left|r_{\Delta\tau}(\tau)\right|\right]}$}.
Clear evidence of the nonstationarity of returns in the physical time
scale is given by the characteristic ``U'' shape assumed by the
intraday volatility defined over the
day-by-day ensemble of returns~\cite{Bassler2007,Baldovin2015,Admati1988,Andersen1997} (see
Fig.~\ref{fig_TimeComparison}a). 
Once analyzed in terms of FST, volatility stationarity is instead 
sensibly recovered (Fig.~\ref{fig_TimeComparison}b).
The availability of a time series with stationary increments,
eliminating thus the seasonalities appearing in the physical time scale,
conveys the methodological advantage that empirical analyses can be
performed through ordinary, sliding-window techniques, hence extending
considerably the sample size for statistical confidence. 

\begin{figure*}[tbp]
\centering \includegraphics[width=0.84 \textwidth]{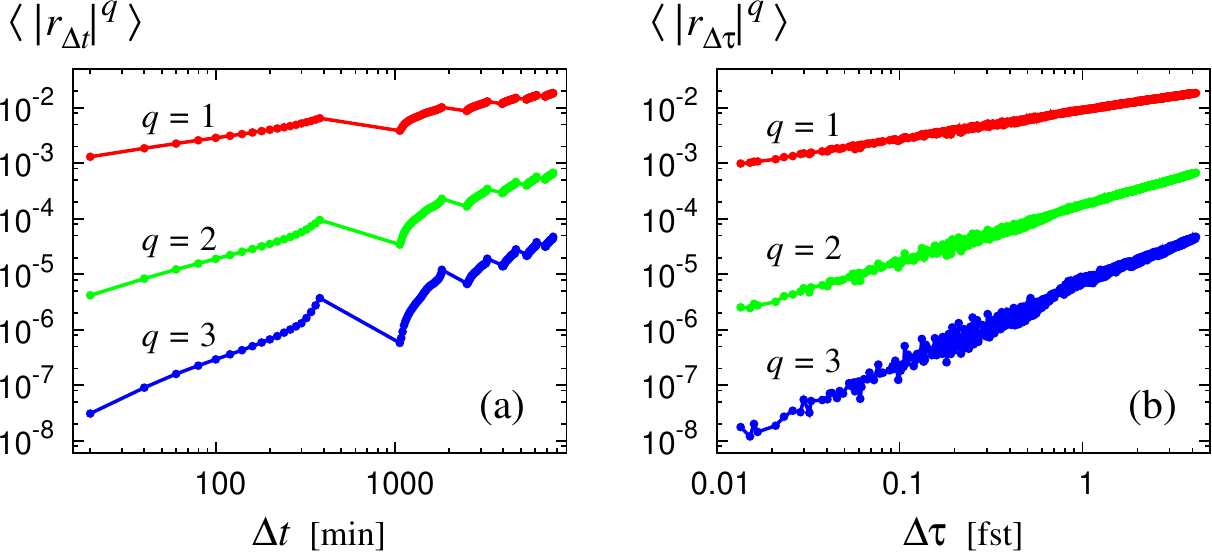}
\caption{
  Empirical moment-analysis. $\Delta t$ (a) and 
  $\Delta\tau$ (b) are the duration of the time intervals
  over which returns $r_{\Delta t}$ and $r_{\Delta\tau}$ are defined,
  respectively. 
}
\label{fig_Scaling}
\end{figure*}

Let us now point out that from  
Eq.~\eqref{eq_scaling2} straightforwardly descends 
$\mathbb{E}\left[|r_{\Delta\tau}|^q\right]=(\Delta\tau)^{q/2}
\,\int\mathrm{d}x\,g(x)\,|x|^q$, or, in the presence of a general Hurst
exponent $H$ like in Eq.~\eqref{eq_scaling}, 
\begin{equation}
\mathbb{E}\left[|r_{\Delta \tau}|^q\right]=(\Delta \tau)^{qH}
\,\int\mathrm{d}x\,g(x)\,|x|^q.
\label{eq_moments}
\end{equation}
One of the consequence of Eq.~\eqref{eq_moments} is that a log-log
plot of the $q$th-moment $\mathbb{E}\left[|r_{\Delta \tau}|^q\right]$
vs. $\Delta \tau$ should be a straight line with slope $qH$. 
The empirical analysis of the returns' moment as a function of the
time interval duration is particularly revealing about the meaning 
of the FST. 
Fig.~\ref{fig_Scaling} highlights that the recurrent patterns
deviating from straight behavior in physical time
(Fig.~\ref{fig_Scaling}a) are wiped out in the FST
(Fig.~\ref{fig_Scaling}b).

\begin{figure*}[tbp]
\centering \includegraphics[width=0.86 \textwidth]{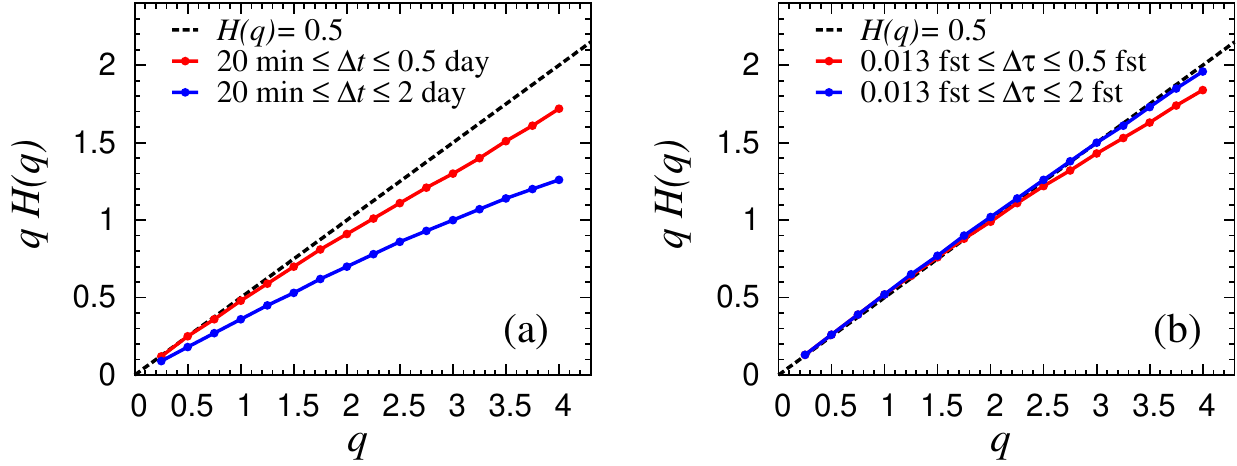}
\caption{
  Multiscaling analysis. Slopes of linear regression of the log-log
  plots in Fig.~\ref{fig_Scaling}a,b over different ranges of the time
  interval are respectively reported as a function of the moment-order
  $q$.
}
\label{fig_Multiscaling}
\end{figure*}

The presence of multiscaling may be easily detected by analyzing the
slopes of the log-log plots in Fig.~\ref{fig_Scaling}. 
As expected, multiscaling features clearly detectable in physical time 
(Fig.~\ref{fig_Multiscaling}a) are still present but become less 
pronounced when the FST is adopted (Fig.~\ref{fig_Multiscaling}b).

\begin{figure*}[tbp]
\centering \includegraphics[width=0.86 \textwidth]{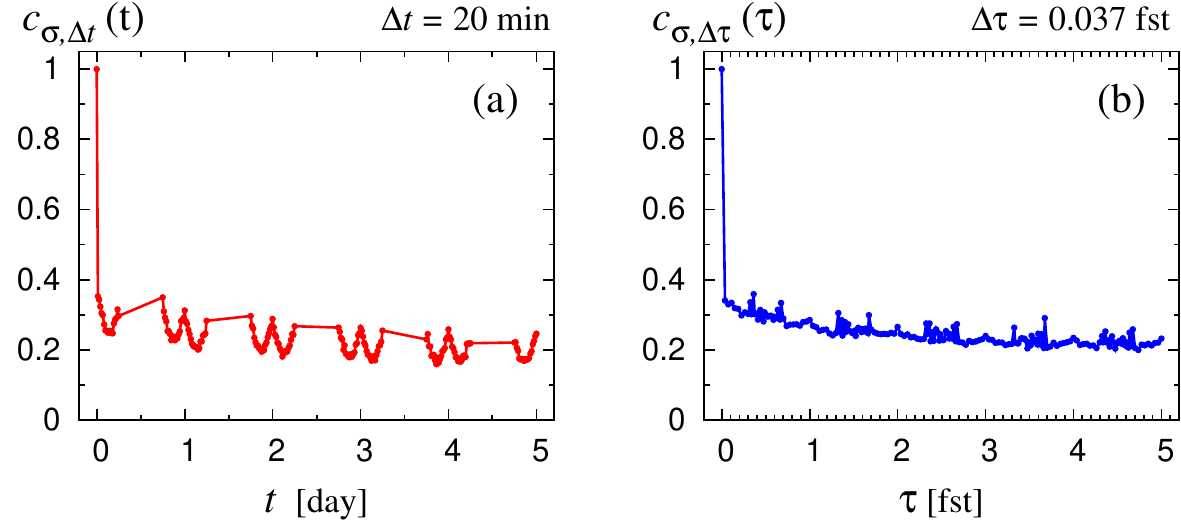}
\caption{
  Volatility autocorrelation as a function of the physical (a) and FST (b) lag. 
  Plots have been realized taking into account all possible pairs of
  lagged returns in the dataset, with $t$ $(\tau)$ an integer multiple 
  of $\Delta t$ ($\Delta\tau$).
  The ciclostationary character of the autocorrelation in physical
  time (a) is suppressed by the adoption of FST (b).
}
\label{fig_Vol_Autocorrelation}
\end{figure*}

So far, results concern single-point statistics. In Finance, a fundamental
two-point indicator is the volatility autocorrelation 
function~\cite{RamaCont2001,Bouchaud2003,Dacorogna2001,Stanley2000},
i.e. the correlation between absolute values of two returns at a given
lag. 
At variance with the linear one, the volatility autocorrelation is
known to decay very slowly with the time lag~\cite{RamaCont2001,Bouchaud2003,Dacorogna2001,Stanley2000}. 
In Fig.~\ref{fig_Vol_Autocorrelation}a 
the $\Delta t=20\;{\rm min}$ volatility autocorrelation at lag $t$,
$c_{\sigma,\Delta t}(t)$~\cite{Volatility_def}, is plotted; 
while Fig.~\ref{fig_Vol_Autocorrelation}b 
reports the analogous quantity in FST, 
$c_{\sigma,\Delta\tau}(\tau)$, for $\Delta \tau=0.037\;{\rm fst}$.
In both cases, the autocorrelation has been evaluated through a
sliding-window procedure. 
The ciclostationary quality of $c_{\sigma,\Delta t}(t)$
in physical time is a direct consequence of the ``U'' 
shape in Fig.~\ref{fig_TimeComparison}a.  
In contrast, in FST-lag the periodic structure of 
$c_{\sigma,\Delta\tau}(\tau)$ is almost completely removed,
pointing out the methodological advantage in such an
empirical estimation.

\section{Conclusions}
We have shown that the adoption as construction criteria of:
(i) simple-scaling invariance with Hurst exponent
$1/2$ for the return PDF, Eq.~\eqref{eq_scaling2};
(ii) validity of the measure property, Eq.\eqref{eq_measure};
leads to a unique time scale for martingale
processes in Finance, which renders stationary the PDFs of returns over
identical spans. 
The construction of FST is based on the minimization of the 
KS distance of the PDF for the returns at a given scale,
with respect to a reference one. 
It covers a wide range 
of physical time scales and opens interesting perspectives in the 
statistical analysis of financial time series. 
For instance, the attained increments stationarity
may sensibly enrich 
empirical estimates enlarging dataset analyses from
ciclostationary to sliding-window procedures. 
Moreover, we have solved the problem of including overnight and over-weekend returns in 
financial analysis, bridging intra and interday regimes.
This makes markets with discontinuous trading activity 
closer to those not experiencing transaction interruptions~\cite{Ghashghaie1996},
opening thus possibilities for meaningful comparisons.

Methodological approaches based on symmetries are not frequently
discussed in Finance. 
However, some of the ingredients we employed have interesting
precursors in the history of financial markets analysis. 
Indeed, stimulated by the discovery of fat tails~\cite{RamaCont2001,Stanley1995,Mandelbrot1963},
previous studies~\cite{Clark1973,Mandelbrot1973,Ane2000}
attempted at modifying the returns PDFs through the introduction of a
stochastic redefinition of time, in order to cast them into Gaussians. 
Here, no reference
to probability models is made and our FST is not stochastic, 
but like in these earlier studies our
time redefinition is intended to alter some fundamental
property of the returns PDFs: in
our case, it enforces a simple scaling symmetry.
While the distribution of geometric Brownian motion of the 
standard model of Finance, or the Levy-stable distribution first 
suggested by Mandelbrot as an alternative~\cite{Mandelbrot1963}, are 
invariant under time rescaling, more recent work on financial
modeling has focused on multiscaling aspects and 
their evocative analogies with
turbulence~\cite{Borland2005,Frisch1997,Bacry2008,Mandelbrot1998}.

Along these lines a multifractal operational time has been assumed
in the modelization of market evolution and multiscaling has been
established as a solid stylized fact in the analysis of series where
the interruption are absent or disregarded~\cite{Bacry2008, Mandelbrot1998}.
One should also mention that
the symmetry property expressed by Eq.~\eqref{eq_scaling}
has been recently assumed as a main modeling ingredient for
the stochastic dynamics of financial
indexes~\cite{Baldovin2007,Stella2010,Peirano2012,Zamparo2013,Baldovin2015a}. 
Combining scaling
with ideas of fine graining inspired by the renormalization group
approach of statistical mechanics~\cite{Kadanoff2005}, these models reproduce
many stylized facts
exposed by interday time series analysis, including multiscaling.

As we have shown here our simple scaling ansatz,
besides trying to satisfy the exigence of a unique time definition,
does not eliminate multiscaling but offers interesting practical advantages.
Among these, as discussed in the Methods section with reference to
Table~\ref{tab_DeltaTau}, is the fact that enforcement of simple scaling
appears to result in a better control of stationarity of the whole
PDF of returns compared to time definitions related to a single moment.

Our construction, which does not amount to the proposal of a specific
parameter dependent model of market dynamics,
does not conflict with the possible presence of other stylized facts
related to the breaking of time reversal invariance,
like the leverage effect~\cite{Bouchaud2001} whereby a negative price
change is on average followed by a volatility increase. 
Indeed, the FST scale-invariant distribution
could, e.g., present skewness.

Time scales capable to conform complex behavior with  
relatively simple properties shared by ordinary stochastic processes 
should also be relevant in the study of
collective assets dynamics,
where time-related issues like e.g. the Epps effect~\cite{Epps1979,Reno2003,Borghesi2007}
or the lead-lag~\cite{Huth2014,Curme2015} question are a main
focus of the current research.
We believe that strategies similar to the one 
adopted here could be used in even broader contexts, in cases in which 
time series display recurrent patterns and scaling properties are 
at least approximately obeyed. One relevant candidate
could be represented by the records of recurrence
times~\cite{Chicheportiche2014,Chicheportiche2017}
either in Finance~\cite{Bogachev2007} or
in seismicity~\cite{Corral2004,Corral2006}.

\section{Methods}
Our FST construction is illustrated considering the S\&P500 index
recorded in one minute intervals between 9:40 am and 16:00 pm New
York time and from 30 September 1985 to 28 June 2013~\cite{DataSetStatement}.
After excluding those days for which the records 
are not complete (e.g. holidays and market anticipated closures or 
delayed openings), the data-set includes $L=6852$ trading days.
For each single day $l$ ($l=1,\ldots,L$) there is a total of $381$ 
index values $s(t_{l,n})$ ($n=0,\ldots,N$, with $N=380$).
In order to work with  zero empirical averages, returns are
detrended: 
$r_{l,n}^{l',n'} \mapsto r_{l,n}^{l',n'}
- \langle r_{l,n}^{l',n'}\rangle $, 
where the average is done over all possible
returns with the same $n$, $n'$ and same $\Delta l=l'-l$,
with $t_{l',n'}>t_{l,n}$.
Overnights include single night returns, returns over weekends, returns 
over holidays and other market closures.
Closures due to specific market reasons 
should be 
in principle treated differently from
weekends and normal holidays
(e.g., several 
days following the black Monday 1987 are missing in the dataset 
because of anticipated market close).
Since we verified that they introduce only minor effects (see also below), 
in our analysis we do not make such distinctions.

In the construction of $\tau_{l,m}$ a key step
consists in checking whether two random variables $X$ and $Y$ are
identically distributed.  
A basic tool to establish this is the KS 
two-sample test~\cite{DeGroot2010,Darling1957}. Given two samples 
$\left\{x_i\right\}_{i=1}^{n_x}$, $\left\{y_i\right\}_{i=1}^{n_y}$, 
with empirical cumulative distribution functions 
$F_x$, $F_y$, respectively,
the test is based on the rescaled supremum of the distributions difference: 
\begin{equation}
D_{x,y}
\equiv \left(\frac{n_x\;n_y}{n_x+n_y}\right)^{1/2}
\;\sup_{z\in\mathbb R}\left\{F_x(z)-F_y(z)\right\}\, .
\label{eq_KStest}
\end{equation}
In those cases in which the samples refer to independent variables,
the test can rely on the 
KS theorem to assign a significance level
to any $D_{x,y}$, under the null hypothesis that
$\left\{x_i\right\}$, $\left\{y_i\right\}$
come from the same distribution.
If however samples are extracted from the same time series 
and data display long-range dependence 
like in
the case examined in the present study,
the mapping between 
$D_{x,y}$ and the associated significance level
is expected to be strongly affected~\cite{Chicheportiche2011},
and only model-dependent statements can be produced.
Still, $D_{x,y}$ quantifies how close the distributions
$F_x$ and $F_y$ are, and here we just employ its minimization
in order to determine $\tau_{l,m}$ within a variational strategy.

\begin{table}[ht]
  \caption{\label{tab_KS1} Value of $D_{x,y}$ between different pairs of samples.} 
\begin{ruledtabular}
\begin{tabular}{c|c|c|c|c|c|c|c}
\multicolumn{8}{c}{\vspace{0.8cm}} \\
 & \begin{rotate}{60} 1 night \end{rotate} & 
\begin{rotate}{60} 3 nights \end{rotate} & 
\begin{rotate}{60} morning \end{rotate} & 
\begin{rotate}{60} afternoon \end{rotate} & 
\begin{rotate}{60} trading day \end{rotate} & 
\begin{rotate}{60} 1 day \end{rotate} & 
\begin{rotate}{60} 2 days \end{rotate} \\ 
\hline 
first 20 min & 10.5 & 6.54 & 13.8 & 12.2 & 17.2 & 17.7 & 18.3 \\ 
\hline 
1 night & • & 0.82  & 3.08 & 2.06 & 6.58 & 8.07 & 10.9 \\ 
\hline 
3 nights & • & • & 1.77 & 1.74 & 3.83 & 5.25 & 6.96 \\ 
\hline 
morning & • & • & • & 1.95 & 4.09 & 6.20 & 9.18 \\ 
\hline 
afternoon & • & • & • & • & 5.85 & 7.95 & 10.4 \\ 
\hline 
trading day & • & • & • & • & • & 2.81 & 6.17 \\ 
\hline 
1 day & • & • & • & • & • & • & 6.96 \\ 
\end{tabular}
\end{ruledtabular}
\end{table}
In Table~\ref{tab_KS1}, we report $D_{x,y}$ calculated 
for different pairs of samples, relative to various intervals in
physical time.
We emphasize the particularly low value of $D_{x,y}$ if calculated
between 1 and and 3 nights. 
As anticipated, this suggests that
it is legitimate to make no distinction between nights and weekends.

\begin{figure}[htbp]
\centering \includegraphics[width=0.5 \textwidth]{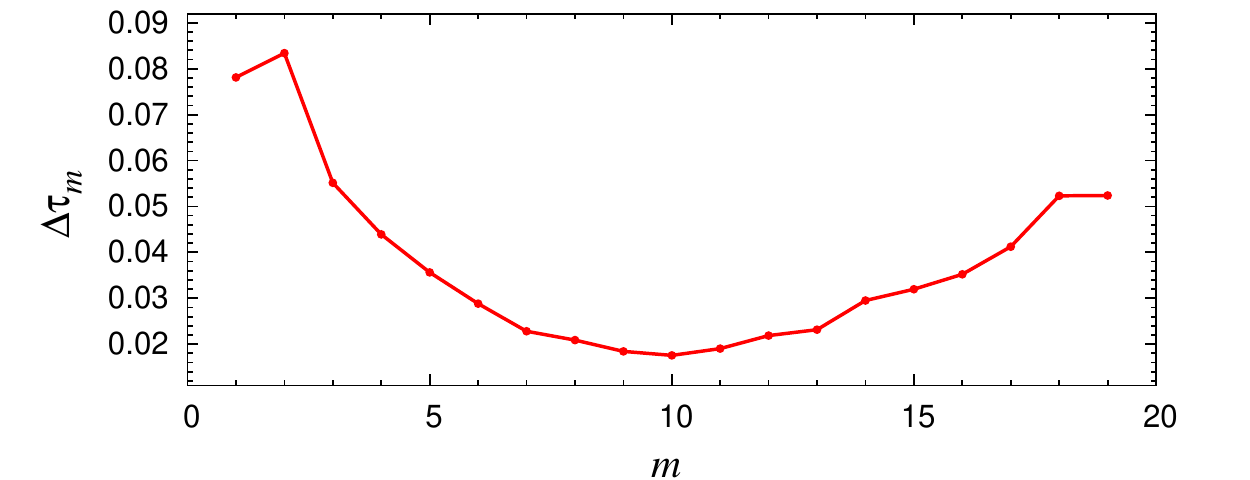}  
\caption{Values of 20-min $\Delta\tau$, $\Delta \tau_m$ ($m=1,\ldots,19$), from day opening to closure.}
\label{fig_DeltaTauInTheDay}
\end{figure}

\begin{table}[ht]
\caption{\label{tab_DeltaTau}
Outcome of the minimization procedure when the one day open-to-open
return distribution is taken as $x$-sample ($n_x=6601$, 
$\Delta \tau_x = 1\;{\rm fst}$).
For $y=10,\;20,\;38\;{\rm min}$,
the last two columns correspond in fact to the average of 
$\Delta \tau_y $ and $D_{x,y}$ over the $(380\;{\rm min}/y)$ 
contiguous time intervals existing within a trading day.
}
\begin{ruledtabular}
\begin{tabular}{c||cccc}
$y$ & $n_y$ & $\Delta \tau_y\;[{\rm fst}] $ & $D_{x,y}$ \\
\hline
10-min & $L$ & $0.011$ & $1.50$  \\
20-min & $L$ & $0.026$ & $1.06$  \\
38-min & $L$ & $0.054$ & $0.87$  \\
morning & $L$ & $0.336$ & $0.70$  \\
afternoon & $L$ & $0.264$ & $0.93$  \\
trading day & $L$ & $0.682$ & $0.64$ \\
overnights & $6601$ & $0.213$ & $0.82$  \\
2 days & $6350$ & $2.150$ & $0.61$  \\
3 days & $3235$ & $3.308$ & $0.79$  \\
5 days & $1187$ & $5.633$ & $0.75$  \\
10 days & $527$ & $10.86$ & $1.11$  \\
\end{tabular}
\end{ruledtabular}
\end{table}
Table~\ref{tab_DeltaTau} displays the minimal value 
of $D_{x,y}$ corresponding to the reported value 
of the FST $\Delta \tau_y$, taking as $x$-sample the open-to-open
return distribution.
While entries for $10$, $20$, $38$ min in Table~\ref{tab_DeltaTau}  
refer to averages over the whole day of intervals of the respective
duration,  
Fig.~\ref{fig_DeltaTauInTheDay} details the values of 20-min $\Delta\tau$ from day opening to closure.

Table~\ref{tab_DeltaTau} clarifies 
the time-window width where Eq.~(\ref{eq_scaling2}) can be
assumed to hold.
The $10\;{\rm min}$ interval is the only case with $D_{x,y}$
significantly above $1$.
To give a comparison, if in place of the 1-day distribution we take a
zero-average Gaussian, the KS analysis
returns $D_{x,y}$ values
between $1.8$ and $4.2$, with the lower values corresponding to the larger
time intervals $y$ as one would expect
in accordance with the slow crossover to
Gaussianity due to the progressive loss of dependence~\cite{RamaCont2001}.
It is also worth mentioning that if one computes $D_{x,y}$ rescaling two samples on the basis of their variance or of any specific moment of the PDF, a larger value of $D_{x,y}$ is obtained if compared with the one guaranteed by the KS procedure.
Correspondingly, a time scale defined on a single-moment rescaling differs up to $25\%$ with respect to the FST (see Table III in Appendix B). In summary, referring to a single moment for the definition of time is a simpler but less efficient procedure to enforce the stationarity of the whole PDF, since not all information at all price-variation scales is employed.

\begin{figure}
\centering \includegraphics[width=0.5 \textwidth]{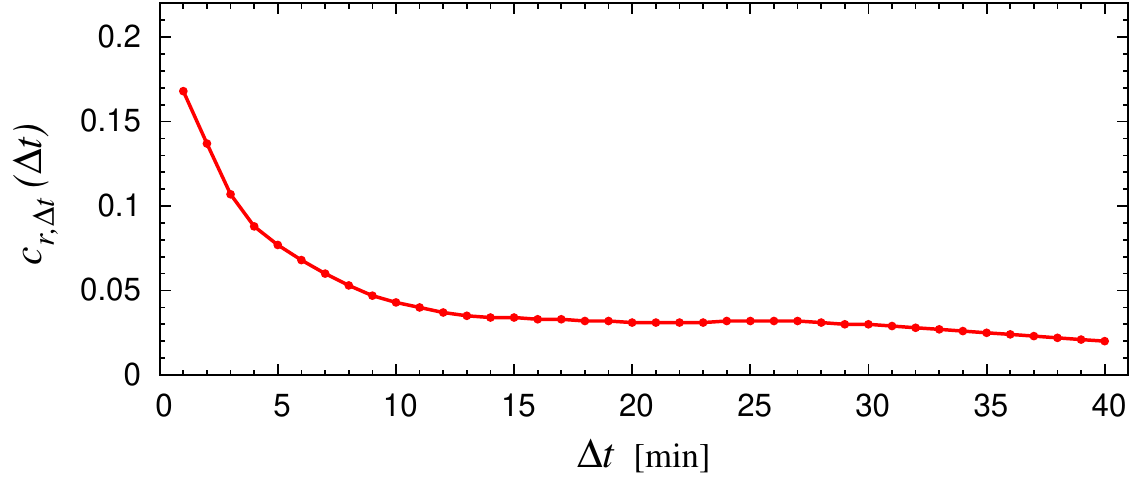}  
\caption{Empirical linear correlation function between two contiguous
  intervals of span $\Delta t$.}
\label{fig_Correlations}
\end{figure}

The $10$ min interval is excluded from the FST window not only in view
of the large $D_{x,y}$,
but also from an analysis of the linear correlations.
Let us indicate with $[\Delta t]\equiv\Delta t/(1\;{\rm min})$ the
number of minutes characterizing $\Delta t$ (for instance, $[10\;{\rm min}]=10$).
The empirical linear correlation between two contiguous intervals of span $\Delta t$ 
may be calculated as 
\begin{eqnarray}
\label{eq_linearcorrelation}
c_{r,\Delta t}(\Delta t) 
&\equiv&  \frac{1}{{N-2[\Delta t] +1}} 
\sum_{n=[\Delta t]}^{N-[\Delta t]} \cdot\\
&&\cdot
\dfrac{\frac{1}{L} \sum_{l=1}^L
r^{l,n}_{l,n-[\Delta t]} \, r^{l,n+[\Delta t]}_{l,n}} {
  \sqrt{\frac{1}{L} \sum_{l=1}^L \left(r^{l,n}_{l,n-[\Delta t]}\right)^2} \,
  \sqrt{\frac{1}{L} \sum_{l=1}^L \left(r^{l,n+[\Delta t]}_{l,n}\right)^2}} \; ,
\nonumber
\end{eqnarray}
and it is plotted in Fig.~\ref{fig_Correlations}.
Linear correlations decrease below our subjective threshold $0.05$ 
only for $\Delta t > 10$ min. Correspondingly, we expect 
the additive property descending from requirement (ii)  to be valid
only for intervals duration above such lower bound.
With the choice $\overline{\Delta t}=20\;{\rm min}$, we verified that 
the difference between the measured scaling factor and the one obtained by the
additivity rule is always positive (consistently with the positive
value of linear correlations) and does  not exceed $25$\% of the 
value of the scaling factor. 
For instance, 
$\Delta \tau_{\mbox{\scriptsize{1\;day}}}
/(\Delta \tau_{\mbox{\scriptsize{morning}}} 
+\Delta \tau_{\mbox{\scriptsize{afternoon}}}
+\Delta \tau_{\mbox{\scriptsize{night}}}) 
\simeq 1.23$.

\section*{Appendix A}

Here we discuss the limitations posed by multiscaling and 
simple scaling with Hurst exponent $H \neq 1/2$ to the possibility
of defining a univoque time scale able to stationarize the PDF
of returns and obeying natural additivity properties.

In the presence of multiscaling, $q$-order moments must
obey the general expression
\begin{equation}
\mathbb{E} \left[ |r|^{q} \right] = A_q \Delta t^{qH(q)} \; ,
\label{eq_multiscalingscaling}
\end{equation}
where $r$ are the returns in an interval of duration $\Delta t$,
$A_q$ is a generic $q$-dependent amplitude and
also the Hurst exponent becomes a function of $q$.


A natural strategy towards stationarization of the returns
over equal time span is that of defining a time scale based
on the choice of a particular moment-order $q$.
If we indicate by $r$ the returns in an interval of
duration $\Delta t$ and by $r_0$ the returns in an interval of
duration $\Delta t_0$ (in our work we chose $\Delta t_0 =1$ day), 
the new time duration $\Delta \tau (q)$ of a generic interval
$\Delta t$ can be defined as
\begin{equation}
\Delta \tau (q) \equiv \dfrac{ \mathbb{E} \left[ |r|^{q} \right]^{\frac{2}{q}} }{\mathbb{E} \left[ |r_0|^{q} \right]^{\frac{2}{q}}} \; .
\label{eq_tau_di_q}
\end{equation}
Notice that in the case of a Wiener process, for which
Eq.~(\ref{eq_scaling2}) is strictly valid with a Gaussian $g$,
the new time does not depend on $q$ and
is equivalent to the physical time.

However, from Eq.~(\ref{eq_multiscalingscaling}) and~(\ref{eq_tau_di_q}) 
one obtains
\begin{equation}
\Delta \tau (q) = \left( \dfrac{\Delta t}{\Delta t_0}\right)^{2H(q)} \; .
\end{equation}
This immediately implies that this time definition is
$q$-independent only if $H(q)$ is constant (no multiscaling).
One could easily prove also the converse argument. Namely, that no
$q$-independent time redefinition can completely eliminate
the existence of nontrivial multiscaling effects.

Furthermore the
time defined by Eq.~(\ref{eq_tau_di_q}) is not additive unless $H = 1/2$.
Indeed, consider $r_1$ and $r_2$ as the returns of two successive contiguous 
intervals $\Delta t_1$ and $\Delta t_2$ with $\Delta t = \Delta t_1 + \Delta t_2$
and $r=r_1+r_2$ as their aggregate, then, with $H \neq 1/2$ we have
\begin{equation}
\Delta \tau_1 (q) + \Delta \tau_2 (q) \neq \Delta \tau (q) \; ,
\end{equation}
where $\Delta \tau$, $\Delta \tau_1$ and $\Delta \tau_2$ are the new time
duration respectively associated to the PDF of $r$, $r_1$ and $r_2$.

The above considerations are at the basis of our ansatz scheme
which assumes the validity of Eq.~(\ref{eq_scaling2}) as the starting
point to construct FST as an additive time independent of
moment order.
\medskip

\section*{Appendix B}

To better clarify the differences between fst and a time scale constructed
starting from Eq.~(\ref{eq_tau_di_q}), we report in Table~\ref{tab_DeltaTau2}
the outcome of different moment $q$ choices.

\begin{table}[ht]
\caption{\label{tab_DeltaTau2}
Outcome of a time definition procedure when the one day open-to-open
return distribution is taken as $x$-sample.
For $y=10,\;20,\;38\;{\rm min}$,
the last two columns correspond in fact to the average of 
$\Delta \tau_y $ and $D_{x,y}$ over the $(380\;{\rm min}/y)$ 
contiguous time intervals existing within a trading day.
Columns with ``fst'' title report data already presented in Table~\ref{tab_DeltaTau} coming from the KS minimization procedure while the other colums are obtained for a time definition based on a single moment $q$; see Eq.~(\ref{eq_tau_di_q}).
}
\begin{ruledtabular}
\begin{tabular}{c||cc|cc|cc|cc}
    &  \multicolumn{2}{c|}{fst} & \multicolumn{2}{c|}{$q=1$} & \multicolumn{2}{c|}{$q=2$} & \multicolumn{2}{c}{$q=3$}  \\
\hline    
    $y$ & $\Delta \tau_y $ & $D_{x,y}$ & $\Delta \tau_y $ & $D_{x,y}$ & $\Delta \tau_y $ & $D_{x,y}$ & $\Delta \tau_y $ & $D_{x,y}$  \\
\hline
10-min       & $0.011$ & $1.50$    & $0.013$ & $2.26$     & $0.015$ & $2.83$    & $0.014$ & $2.77$ \\
20-min       & $0.026$ & $1.06$    & $0.031$ & $2.26$     & $0.031$ & $1.90$    & $0.032$ & $2.12$ \\
38-min       & $0.054$ & $0.87$    & $0.066$ & $1.38$     & $0.061$ & $1.52$    & $0.069$ & $1.84$ \\
morning      & $0.336$ & $0.70$    & $0.347$ & $0.81$     & $0.325$ & $0.82$    & $0.264$ & $1.99$ \\
afternoon    & $0.264$ & $0.93$    & $0.305$ & $1.57$     & $0.353$ & $2.38$    & $0.392$ & $3.00$ \\
trading day  & $0.682$ & $0.64$    & $0.707$ & $0.87$     & $0.733$ & $1.06$    & $0.784$ & $1.47$ \\
overnights   & $0.213$ & $0.82$    & $0.238$ & $1.24$     & $0.248$ & $1.41$    & $0.214$ & $0.86$ \\
2 days       & $2.150$ & $0.61$    & $2.074$ & $0.64$     & $2.164$ & $0.62$    & $2.294$ & $0.87$ \\
3 days       & $3.308$ & $0.79$    & $3.033$ & $1.00$     & $2.889$ & $1.09$    & $2.755$ & $1.24$ \\
5 days       & $5.633$ & $0.75$    & $5.186$ & $0.91$     & $4.946$ & $1.00$    & $4.724$ & $1.12$ \\
10 days      & $10.86$ & $1.11$    & $9.318$ & $1.13$     & $9.615$ & $1.13$    & $9.541$ & $1.11$ \\
\end{tabular}
\end{ruledtabular}
\end{table}
Results show that a time definition based on the second moment may differ up to about
25 \% with respect to our fst, defined via the Kolmogorov-Smirnov test and that time scales defined on the other considered moments ($q=1,3$) do not perfom better as far as KS distance is concerned.
Furthermore the Table shows that, to define a time scale which enforces the scaling symmetry Eq.~(\ref{eq_scaling2}), there is not a preferential moment since the best choice would depend on the considered time interval.
For example in the case of overnight returns a time definition based on the third moment would perform much better in the attempt to satisfy Eq. (2).
This shows that if the goal is to promote in the best possible way the stationarity of the return pdf’s (and this is our goal), the KS criterion on which we rely in our procedure is much more adequate than any moment based procedure.

\section*{Acknowledgments}
We acknowledge support from the Research Project ``Dynamical
behavior of complex systems: from scaling symmetries to economic
growth'' of the University of Padova.


\begin{thebibliography}{41}
\bibitem{DataSetStatement}
Whereas New York market opens at 9:30 am, our dataset records start instead at 9:40 am.
\bibitem{Volatility_def}
The volatility autocorrelation is defined as
$
c_{\sigma,\Delta t}(t)\equiv
\frac{\langle\left|r_{\Delta t}(t'+t)\;r_{\Delta t}(t')\right|\rangle_{t'}
  -\langle\left|r_{\Delta t}(t'+t)\right|\rangle_{t'}\;\langle\left|r_{\Delta t}(t')\right|\rangle_{t'}}
     {\sqrt{
       \left(
       \langle\left[r_{\Delta t}(t'+t)\right]^2\rangle_{t'}-\langle\left[r_{\Delta t}(t'+t)\right]\rangle_{t'}^2
       \right)
       \left(
       \langle\left[r_{\Delta t}(t')\right]^2\rangle_{t'}-\langle\left[r_{\Delta t}(t')\right]\rangle_{t'}^2
       \right)
       }},
$
where $\langle\cdot\rangle_{t'}$ is a sliding-window empirical 
average with respect to $t'$, performed over all available couples $(t',t'+t)$ in the dataset.
An analogous definition follows for $c_{\sigma,\Delta\tau}(\tau)$ in FST.
\end{thebibliography}
\end{document}